\documentclass[12pt]{article}

\usepackage{amsmath}
\usepackage{amsfonts}
\usepackage{eufrak}
\usepackage{amssymb,bbold}
\usepackage{epsfig}
\usepackage{bm}

\def\be{\begin{equation}}
\def\ee{\end{equation}}

\def\n{\nabla}
\def\t{\tau}

\def\v{\nu}

\def\m{\mu}
\def\s{\sigma}

\def\o{\omega}

\def\pa{\partial}

\def\e{\epsilon}
\def\G{\Gamma}
\def\md{\mathcal{D}}

\def\d{\delta}

\def\6{\partial}

\def\mbb{\mathbb{R}}

\newcommand{\ip}{\raise1pt\hbox{\large$\lrcorner$}\,}
\newcommand{\bea}{\begin{eqnarray}}
\newcommand{\eea}{\end{eqnarray}}
\newcommand{\nn}{\nonumber \\}

\begin{document}
\title{Classification of supersymmetric spacetimes in eleven dimensions}

\author{Marco Cariglia \thanks{M.Cariglia@damtp.cam.ac.uk} and
  Ois\'{\i}n A. P. Mac
  Conamhna\thanks{O.A.P.MacConamhna@damtp.cam.ac.uk} \\ DAMTP \\ Centre
  for Mathematical Sciences \\ University of Cambridge \\ Wilberforce
  Road, Cambridge CB3 0WA, UK.}

\maketitle 

\abstract{We derive, for spacetimes admitting a Spin(7) structure, the
general local bosonic solution of the Killing spinor equation of
eleven dimensional supergravity. The metric, four form
and Killing 
spinors are determined explicitly, up to an arbitrary eight-manifold
of Spin(7) holonomy. It is sufficient to impose the
Bianchi identity and one particular component of the four form field
equation to ensure that the solution of the Killing spinor equation
also satisfies all the field equations, and we give these conditions
explicitly.} 

\section{Introduction and results}
Supersymmetric spacetimes are of central importance to M- and string
theory. They have played a key role in the discovery, verification and
subsequent generalisation of gauge theory-gravity duality, through
the AdS/CFT correspondence. They have also provided one of the most
fruitful areas of overlap between pure mathematics and theoretical
physics, through studies of special holonomy manifolds, G-structures
and mirror symmetry. The problem of classifying supersymmetric solutions of
supergravity theories has therefore attracted much attention, and
numerous distinct approaches have been employed. It was shown in
\cite{gmpw} that the notion of a G-structure allows for the deduction
of an explicit set of necessary and sufficient algebraic conditions on the spin
connection and fluxes for the existence of a single Killing spinor in
any supergravity,
and this approach has since been applied in numerous contexts, to
classify minimally supersymmetric spacetimes in various supergravity
theories. In particular, this analysis was applied to d=11
supergravity in \cite{gaunt1}, \cite{gaunt}. In \cite{me}, the
standard G-structure formalism was refined to give a universally
applicable formalism for the complete classification of all
supersymmetric spacetimes, admitting any desired number of arbitrary
Killing spinors, in any supergravity. The complete and explicit nature
of this formalism make it an appropriate tool to apply to the
exhaustive study of supersymmetric spacetimes in M- and string
theory. The end result of this study will be a complete catalogue of
the geometry and matter of all such spacetimes. Such a catalogue will
certainly be of relevance to the AdS/CFT correspondence, and will be
likely to produce many more applications.

Thus, an obvious target for the refined formalism is d=11 supergravity. In
\cite{us}, all structure groups arising as subgroups of the isotropy
group of a single null Killing spinor were classified, and the spaces of
spinors fixed by these groups were constructed. These results can now
be used to undertake the refined classification of all supersymmetric
spacetimes in eleven dimensions admitting at least one null Killing
spinor. In this letter we initiate this classification, by deriving the
general local bosonic solution of the Killing spinor equation for two
Killing spinors sharing a common Spin(7) isotropy group.

Very recently, in \cite{pap}, the method of \cite{me} was
reformulated (using more abstract notation), and its application to
the complete 
classification of all supersymmetric eleven-dimensional spacetimes
admitting at least one timelike Killing spinor was initiated.

As was shown in
\cite{us}, the existence of an arbitrary Spin(7) structure in eleven dimensions
is equivalent to the existence of a pair of Killing spinors
\be
\e,\;\;\; (f+u_i\G^i+g\G^-)\e,
\end{equation}
where $f$, $u_i$ and $g$ are ({\it a priori}) arbitrary real functions, $g\neq0$, and the spinor $\e$ satisfies the projections 
\bea
\G_{1234}\e=\G_{3456}\e=\G_{5678}\e=\G_{1357}\e&=&-\e,\nn\G^+\e&=&0,
\eea
in the spacetime basis 
\be
ds^2=2e^+e^-+\d_{ij}e^ie^j+(e^9)^2,
\end{equation}
where $i,j=1,...,8$. The eight-manifold spanned by the $e^i$ will be
referred to as the base. Our result is that assuming the existence
of this pair of Killing spinors, and nothing else, the general bosonic solution of the Killing spinor equation
admitting a Spin(7) structure is determined locally as follows. We may
take the Killing spinors to be $\e$, $H^{-1/3}(x)\G^-\e$, with
metric
\bea
ds^2&=&H^{-2/3}(x)\Big(2[du+\lambda(x)_Mdx^M][dv+\v(x)_Ndx^N]+[dz+\s(x)_Mdx^M]^2\Big)\nn&+&H^{1/3}(x)h_{MN}(x)dx^Mdx^N,
\eea
where $h_{MN}$ is a metric of Spin(7) holonomy and $d\lambda$, $d\v$
and $d\s$ are two-forms in the $\mathbf{21}$ of Spin(7). Observe that
there are three Killing vectors. Defining the
elfbeins 
\bea
e^+&=&H^{-2/3}(du+\lambda),\nn e^-&=&dv+\v,\nn
e^9&=&H^{-1/3}(dz+\s),\nn e^i&=&H^{1/6}\hat{e}^i(x)_Mdx^M,
\eea
where $\hat{e}^i$ are the achtbeins for $h$, the four-form is
\bea
F&=&e^+\wedge e^-\wedge e^9\wedge d\log H+H^{-1/3}e^+\wedge e^-\wedge
d\s-e^+\wedge e^9\wedge d\v\nn&+&H^{-2/3}e^-\wedge e^9\wedge
d\lambda+\frac{1}{4!}F^{\mathbf{27}}_{ijkl}\hat{e}^i\wedge\hat{e}^j\wedge\hat{e}^k\wedge\hat{e}^l.
\eea
Similar flux terms to those arising in our general solution have been
recognised before in different contexts. The $F^{\mathbf{27}}$ term
was used in \cite{pope} 
to construct resolved membrane solutions. Similar fibrations to those
arising here were used to construct rotating and/or wrapped M2 brane
solutions with a Calabi-Yau transverse space in, for example,
\cite{gaunt1}, \cite{lu}. In the case of a Spin(7) structure, the
terms given above exhaust all 
possibilities.   

Typically, the Killing spinor equation provides a first integral of some
(but not all) of the field equations and Bianchi identities. We will
show below that it is sufficient to impose the Bianchi 
identity and the $+-9$ component of the four-form field equation
$\star(d\star F+\frac{1}{2}F\wedge F)=0$ on our solution of the
Killing spinor equation to
ensure that all field equations are satisfied. The Bianchi identity
reduces to
\be\label{1}
dF^{\mathbf{27}}=0,
\end{equation} 
which implies that $F^{\mathbf{27}}=F^{\mathbf{27}}(x)$, and
$\tilde{d}F^{\mathbf{27}}=0$, where $\tilde{d}$ is the exterior
derivative restricted to the base. The field equation is
\be\label{2}
\tilde{\n}^2H=-\frac{1}{2}d\s_{MN}d\s^{MN}-d\lambda_{MN}d\v^{MN}-\frac{1}{2\times4!}F^{\mathbf{27}}_{MNPQ}F^{\mathbf{27}MNPQ},
\end{equation}
where $\tilde{\n}^2$ is the Laplacian on the eight-manifold with
metric $h_{MN}$, and in this equation all indices are raised with $h^{MN}$.
Our general solution
of the Killing spinor equation is determined explicitly, up to an arbitrary eight-manifold with Spin(7)
holonomy. We have nothing new to say about classifying Spin(7)
holonomy manifolds here. Rather, we regard eight-manifolds of Spin(7)
holonomy as the input for our 
results, with the output, for a particular Spin(7) manifold, being the
general $\mathcal{N}=2$ solution of the Killing spinor equation induced by that
manifold. More precisely, our results, together with those of
\cite{gaunt}, may usefully be thought of as
explicitly providing the most general map from the space of
eight-manifolds with Spin(7) holonomy to the space of solutions of the
Killing spinor equation in eleven dimensions. We expect that
qualitatively similar results will be 
obtained for all other G-structures in eleven dimensions.

\section{Deriving the solution}
Let us briefly discuss the method we use to derive the solution; more
details may be found in \cite{me}, \cite{us}. We assume that the null
spinor $\e$ is Killing. The constraints associated with its existence
were derived in \cite{gaunt}. A useful set of
projections satisfied by $\e$, together with miscellaneous definitions
and identities for Spin(7) forms which we use throughout, are given in
\cite{gaunt}. We adopt all the conventions and notation of this paper,
so that we may readily incorporate their results.  We may contruct a basis for spinor space
by acting on $\e$ with a subset of the Clifford
algebra; it was shown in \cite{us} that a basis is given by the
thirty-two spinors
\be\label{nice}
\e,\;\;\;\G^i\e,\;\;\;J^A_{ij}\G^{ij}\e,\;\;\;\G^-\e,\;\;\;\G^{-i}\e,\;\;\;J^A_{ij}\G^{-ij}\e,
\end{equation}
where the forms $J^A_{ij}$, $A=1,...,7$ furnish a basis for the
$\mathbf{7}$ of Spin(7). The most general additional Killing
spinor compatible with the existence of a Spin(7) structure is
$\eta=(f+u_i\G^i+g\G^-)\e$, $g\neq0$. We may simplify $\eta$
by acting on it with an element of the isotropy group of $\e$, thus
leaving the constraints on the intrinsic torsion derived in
\cite{gaunt} invariant. By acting with with the
$(Spin(7)\ltimes\mbb^8)\times\mbb$ element
\be
1+g^{-1}u_i\G^{+i}+g^{-1}f\G^{+9},
\end{equation}
we may always set $\eta=g\G^-\e$, which we do in what follows.
Now, since $\e$ is Killing, $g\G^-\e$ is Killing if and
only if 
\be
[\md_{\m},g\G^-]\e=0,
\end{equation}
where $\md_{\m}$ is the supercovariant derivative. We may impose the
projections satisfied by $\e$ to write each spacetime component of
$[\md_{\m},g\G^-]\e$ as a manifest sum of the basis spinors
(\ref{nice}). Then by the linear independence of the basis, the
coefficient of each basis spinor in each spacetime component must
vanish separately. 

\subsection{Calculating $[\md_{\m}, g\G^-]\e$.}
Let us evaluate 
\be\label{spin}
[\md_{\m}, g\G^-]\e,
\end{equation}
using throughout the results of \cite{gaunt}, and using the algebraic
relationships derived in that paper to express the
components of the four form in terms of the spin
connection, wherever possible. The $+$ component of (\ref{spin}) is 
\be
\Big[-g\o_{++9}-g\o_{++i}\G^i+\pa_+g\G^-+\frac{g}{3}\o^{\mathbf{7}}_{+ij}\G^{-ij}-\frac{2g}{3}\o_{+9i}\G^{-i}\Big]\e.
\end{equation}
The $-$ component is
\be
\Big[g(\o_{+9-}-\o_{-+9})+\frac{g}{3}(2\o_{99i}-\o_{i-+}-3\o_{-+i})\G^i+\frac{2g}{3}\o^{\mathbf{7}}_{ij9}\G^{ij}+\pa_-g\G^-\Big]\e.
\end{equation}
The $9$ component is
\be
\Big[-g\o_{9+9}-\frac{g}{3}(3\o_{9+i}+\o_{+9i})\G^i+\frac{g}{3}\o_{+ij}^{\mathbf{7}}\G^{ij}+\pa_9g\G^--\frac{2g}{3}(\o_{99i}+\o_{i-+})\G^{-i}-\frac{g}{3}\o^{\mathbf{7}}_{ij9}\G^{-ij}\Big]\e.
\end{equation}
The $i$ component is
\bea
&&\Big[g(-\o_{i+9}+\frac{1}{3}\o_{+9i})+g(\o_{ij+}-\frac{1}{3}\o^{\mathbf{7}}_{+ij}+\frac{1}{2}F^{\mathbf{21}}_{+9ij})\G^j\nn&+&g(-\frac{2}{21}g_{ij}\o_{+9k}+\frac{1}{4}F^{\mathbf{48}}_{+ijk})\G^{jk}
  +(\pa_ig+\frac{g}{3}(2\o_{i-+}-\o_{99i}))\G^-\\&+&g(-\frac{1}{2}g_{ij}\o_{+-9}+\frac{4}{3}\o_{ij9}^{\mathbf{7}}+\o_{ij9}^{\mathbf{35}})\G^{-j}+g(\frac{2}{21}g_{ij}(\o_{99k}+\o_{k-+})+3(\o^{\mathbf{7}}_{ijk})^{\mathbf{48}})\G^{-jk}\Big]\e.\nonumber
\eea

\subsection{Contraints for a Spin(7) structure}
Now we impose
\be
[\md_{\m}, g\G^-]\e=0.
\end{equation}
Setting the coefficient of
each basis spinor in each spacetime component of (\ref{spin}) to zero,
and using the $N=1$ constraints of \cite{gaunt}, we find
that the only non-zero components of the spin connection are
\bea
\o_{+-i}=\o_{-+i}=\o_{99i}&=&\o_{i+-}=\pa_i\log g,\nn
\o_{ij+}^{\mathbf{21}}&=&-\frac{1}{2}F^{\mathbf{21}}_{+9ij},\nn\o_{ij-}^{\mathbf{21}}&=&\frac{1}{2}F^{\mathbf{21}}_{-9ij},\nn\o_{ij9}^{\mathbf{21}}&=&\frac{1}{2}F^{\mathbf{21}}_{+-ij},\nn\o_{ijk}&=&-\frac{1}{4}\d_{i[j}\pa_{k]}\log
g+\frac{1}{8}\phi_{ijk}^{\;\;\;\;\;l}\pa_l\log
g+\o_{ijk}^{\mathbf{21}},
\eea
together with $\o_{+jk}^{\mathbf{21}}$, $\o_{-jk}^{\mathbf{21}}$ and
$\o_{9jk}^{\mathbf{21}}$, which drop out of the Killing spinor
equations for $\e$ and $g\G^-\e$ and are unconstrained. In the above,
$\o_{ijk}^{\mathbf{21}}$ denotes the $\mathbf{21}$ projection of
$\o_{ijk}$ on $j$, $k$. The four-form is required to be
\bea
F&=&-3e^+\wedge e^-\wedge e^9\wedge d\log g+e^+\wedge
e^-\wedge(\o_{ij9}^{\mathbf{21}}e^i\wedge e^j)-e^+\wedge
e^9\wedge(\o_{ij+}^{\mathbf{21}}e^i\wedge e^j)\nn&+&e^-\wedge
e^9\wedge(\o_{ij-}^{\mathbf{21}}e^i\wedge
e^j)+\frac{1}{4!}F^{\mathbf{27}}_{ijkl}e^i\wedge e^j\wedge e^k\wedge
e^l.
\eea
$F^{\mathbf{27}}_{ijkl}$ drops out of the Killing spinor equations for
$\e$ and $g\G^-\e$ and is unconstrained. The function $g$ is required to satisfy
\be\label{dg}
\pa_+g=\pa_-g=\pa_9g=0.
\end{equation}         

\subsection{Solving the constraints}
Given our constraints on the spin connection, by repeating the arguments of \cite{gaunt}, we see that we may
consistently introduce their local coordinates. Thus we take
\bea
e^+&=&L^{-1}(du+\lambda),\nn
e^-&=&dv+\frac{1}{2}\mathcal{F}du+Bdz+\v,\nn\label{coord} e^9&=&C(dz+\s),\nn
e^i&=&e^i_Mdx^m.
\eea
The one-forms $\lambda$, $\v$ and $\s$ have components only on
the base, and $L$, $\mathcal{F}$, $B$, $C$, $\lambda$, $\v$ $\s$ and
$e^i_M$ are independent of the coordinate $v$. Note that with this
choice of coordinates, (\ref{dg}) implies that 
\be
g=g(x).
\end{equation}
Now, using
$de^{\m}=\o_{\v\s}^{\;\;\;\;\;\m}e^{\v}\wedge e^{\s}$, and employing
our constraints from the previous subsection, we find
\bea
de^+&=&2d\log g\wedge e^++\o_{ij-}^{\mathbf{21}}e^i\wedge e^j,\nn
de^-&=&\o_{ij+}^{\mathbf{21}}e^i\wedge e^j,\nn de^9&=&d\log g\wedge
e^9+\o_{ij9}^{\mathbf{21}}e^i\wedge e^j.
\eea
Comparing these expressions with the exterior derivatives of (\ref{coord}), we find that locally
we may take $L^{-1/2}=C=g$, $\mathcal{F}=B=0$, $\lambda=\lambda(x)$,
$\v=\v(x)$, $\s=\s(x)$, and
$d\lambda,$ $d\v,$ $d\s\in\Lambda^2_{21}$. Next we want to solve the
constraints on the base space. Defining
\bea
M_{ij}&=&\d_{ik}(\pa_ue^k)_j,\nn\Lambda_{ij}&=&\d_{ik}(\pa_ze^k)_j,
\eea
we find on comparing our constraints on the spin connection (given the
form of $e^+$, $e^-$, $e^9$ we have just derived) with the explicit
expressions worked out in Appendix D of \cite{gaunt}, that
\bea
M_{ij}&=&M_{ij}^{\mathbf{21}},\nn\Lambda_{ij}&=&\Lambda_{ij}^{\mathbf{21}},
\eea
and hence that
\bea
\o_{ijk}^{\mathbf{7}}=\tilde{\o}_{ijk}^{\mathbf{7}}&=&-\frac{1}{4}\d_{i[j}\pa_{k]}\log
g+\frac{1}{8}\phi_{ijk}^{\;\;\;\;\;l}\pa_l\log
g,\nn\o_{ijk}^{\mathbf{21}}&=&\tilde{\o}_{ijk}^{\mathbf{21}}+\s_i\Lambda_{jk}+\lambda_i
M_{jk},
\eea
where $\tilde{\o}$ denotes the spin connection of the
base. Conformally rescaling the base according to
$e^i=g^{-1/2}\hat{e}^i$, we find that $\hat{\o}_{ijk}^{\mathbf{7}}=0$,
where $\hat{\o}$ denotes the spin connection of the conformally
rescaled base. Thus the conformally rescaled
metric is a metric of Spin(7) holonomy. In fact, we may take the base
to be independent of $u$ and $z$. To see this, note that we have the
freedom to perform Spin(7) transformations on the base preserving
$\e$, $g\G^-\e$, and thus the intrinsic torsion. Under such a
transformation, the basis transforms as
\be
\hat{e}^i\rightarrow (\hat{e}^i)^{\prime}=Q^i_{\;\;j}\hat{e}^j.
\end{equation}
By performing a $v$-independent Spin(7) transformation, we may choose
a $u$-independent basis $(\hat{e}^i)^{\prime}$ if we can find a Spin(7)
  matrix $Q$ such that 
\be
M=(\pa_uQ^{-1})Q
\end{equation}
Since $M$ is required to be in the adjoint of Spin(7), we may always find such a $Q$
locally. Repeating the argument for a $u$, $v$ independent Spin(7)
transformation we find that we can also take the $\hat{e}^i$ to be
independent of $z$. Finally defining $g=H^{-1/3}$, and expressing the
four-form in terms of $\lambda$, $\v$ and $\s$, we obtain the general
Spin(7) solution
in the form given in the introduction.

\subsection{Integrability conditions}
Let us assume that we impose the four-form Bianchi identity, $dF=0$,
on our solution of the Killing spinor equation. We want to use the
integrability conditions for the Killing spinor equation to determine
which of the field equations we must impose to ensure that all are
satisfied. Given that the Bianchi identity is satisfied, the
(contracted) integrability condition for an arbitrary Killing spinor
$\rho$ is
\be\label{intt}
\G^{\v}[\md_{\m},\md_{\v}]\rho=(E_{\m\v}\G^{\v}+Q_{\v\s\t}\G_{\m}^{\;\;\v\s\t}-6Q_{\m\v\s}\G^{\v\s})\rho=0,
\end{equation}
where $E_{\m\v}=0$ and $Q_{\m\v\s}=0$ are the Einstein and four-form
field equations. Taking $\rho=\e$, by writing each spacetime component
of (\ref{intt}) as a manifest sum of basis spinors, we find the
following algebraic relationships between the components of the field
equations:
\bea
E_{+-}=E_{99}&=&12Q_{+-9},\nn
E_{+i}&=&18Q_{+i9},\nn
E_{ij}&=&-6Q_{+-9}\d_{ij}.
\eea
The components $E_{++}$ and $Q_{+ij}^{\mathbf{21}}$ drop out, and are
unconstrained by the integrability condition for $\e$, but all other
components of the field equations are required to vanish
identically. Next taking $\rho=g\G^-\e$, we find the additional
conditions
\be
E_{++}=E_{+i}=Q_{+i9}=Q_{+ij}^{\mathbf{21}}=0.
\end{equation}
Thus given the existence of the pair of Killing spinors $\e$,
$g\G^-\e$, it is sufficient to impose the four-form Bianchi identity
and the single equation $Q_{+-9}=0$ to ensure that all field equations
are satisfied.

\section{Acknowledgements}
M. C. is supported by EPSRC, the Cambridge European Trust and
Fondazione Angelo Della Riccia. O. C. is supported by a Senior Rouse
Ball Scholarship.

\end{document}